\documentclass[12pt]{article}
\usepackage{epsfig}

\def\beq{\begin{equation}}
\def\eeq#1{\label{#1}\end{equation}}
\def\eeqn{\end{equation}}

\def\beqa{\begin{eqnarray}}
\def\eeqa#1{\label{#1}\end{eqnarray}}
\def\eeqan{\end{eqnarray}}

\let\bar=\overbar

\def\Dslash{\not{\hbox{\kern-4pt $D$}}}
\def\dslash{\not{\hbox{\kern-2pt $\del$}}}

\def\msb{{\bar{\ssstyle M \kern -1pt S}}}

\usepackage{fancyhdr,graphicx}
\def\Title#1{\begin{center} {\Large {\bf #1} } \end{center}}

\begin{document}

\Title{Combustion gasdynamics of the neutron $\rightarrow$ strange
matter conversion: towards an assessment of realistic scenarios}

\bigskip\bigskip


\begin{raggedright}

{\it J.E. Horvath\index{}\\
Instituto de Astronomia, Geof\'\i sica e Ci\^encias Atmosf\'ericas\\
Universidade de S\~ao Paulo\\
Rua do Mat\~ao 1226\\
05508-900 S\~ao Paulo SP\\
Brazil\\
{\tt Email: foton@astro.iag.usp.br}}
\bigskip\bigskip
\end{raggedright}

\section{Abstract}

A variety of descriptions of the conversion of a neutron into a
strange star have appeared in the literature over the years.
Generally speaking, these works treat the process as a mere phase
transition or ignore everything but microscopic kinetics,
attempting to pin down the speed of the conversion and its
consequences. We revisit in this work the propagation of the
hypothetical ``combustion'' $n \rightarrow SQM$ in a dense stellar
environment. We address in detail the instabilities affecting the
flame and present new results of application to the turbulent
regime. The acceleration of the flame, the possible transition to
the distributed regime and further deflagration-to-detonation
mechanism are addressed.  As a general result, we conclude that
the burning happens in (at least) either the turbulent
Rayleigh-Taylor or the distributed regime. In both cases the
velocity of the conversion of the star is several orders of
magnitude larger than $v_{lam}$, making the latter irrelevant in
practice for this problem. A transition to a detonation is by no
means excluded, actually it seems to be favored by the physical
setting, but a definitive answer would need a full numerical
simulation.

\section{Introduction}

The interest in hypothetical absolutely stable phases of QCD at high density
(~\cite{oldies}) has been going on for more than two decades. In addition to the
celebrated and widely studied SQM, renewed work on pairing in dense matter (~\cite{pair})
opened the possibility of an extra gain of energy, possibly enhancing the
prospects for a ``$\Delta$SQM''stability (~\cite{nos1}).

There are many approaches to this problem dealing with evolved configurations
of compact stars made from SQM or $\Delta$SQM (refs) and their observable features.
Another important question is how exactly the ``normal'' (proto) neutron stars
become exotic objects. This could happen early in their lives (on $\sim seconds$
{\it via} classical nucleation or later if the conversion is quenched and
depends ultimately on accretion, driven by quantum effects
(bombaci). Even though the compression of the central regions of the star to
$\rho \geq 3 \rho_{0}$ and temperatures $T \sim 10 MeV$ favor the conversion
$n \rightarrow SQM$,it is still possible that the nucleation could be
postponed ~\cite{IB}. In the remaining of this work we will assume a ``hot'' conversion
on $\sim s$ timescale, focusing on the precise form of propagation and related
energetic issues.

The fate of a just-nucleated stable drop of SQM (or $\Delta$ SQM)
matter has been addressed in several works (for example, ~\cite{angela}, ~\cite{GB}).
Usually,it was {\it assumed} that the propagation of the
conversion front proceeds by diffusion of the s-quarks ahead,
carrying the seed of flavor conversion that releases energy after
relaxation. This ``chemical'' equilibration depends on weak
interactions and occurs in a timescale $\tau_{w} \sim 10^{-9}
s$,much faster than dynamical times of the order of a $ms$. In
some works not even the formation of a front is acknowledged, and
a macroscopic synchronization of the conversion is assumed, that
is, deconfinement followed by a decay of $d$ into $s$ quarks, a
process which is unlikely given that the conditions for that
should be extreme in a region of the order of $\sim 10 km$
(~\cite{hind}, ~\cite{HSS}).

In general, kinetic descriptions postulating diffusion lead to a well-known
slow combustion mode termed {\it deflagration}, which is familiar to anyone. Moreover,
in these works only the {\it laminar} regime is studied,leading to front velocities
$\ll c_{sound} \sim c$,neglecting therefore all
complications related to the behavior actually observed in flames. Specifically,it has
been known for years (~\cite{LD}) that at least two instabilities are important for the
flame propagation: the Landau-Darrehius and Rayleigh-Taylor modes. While the first
desestabilizes all wavelengths at the liner level (but is controlled by non-linear
terms),the latter acts on large scales and is always active and important. Both
instabilities develop on timescales $\sim \tau_{dynamical} \sim 10^{-3} s$, and therefore
their effects should be considered from the scratch, right after the beginning of the
propagation.

\section{The cellular stage of the propagation}

After a few ms, the action of the LD instability leads to a
wrinkling of the flame and the development of a so-called {\it
cellular structure}. As a result of this, the fuel consumption
rate rises (because it is controlled by the total area, which has
become bigger) and the flame accelerates. However, the non-linear
effects neglected by Landau play an important role stabilizing the
flame, which looks nested Several approaches have been attempted
One of these employed concepts of fractal geometry to argue that
the velocity of the flame (by itself a fractal) can be described
by the expression ~\cite{BlinSas}

\begin{equation}
u_{cell} = u_{lam} {\bigg({l_{max}\over{l_{min}}}\bigg)}^{0.6
(1-{\rho_{2}/\rho_{1}})^{2}}
\end{equation}

Whereas the maximum length $l_{max}$ is not difficult to find, and
is ultimately bounded by the largest scale in the flame, the
minimum length is more tricky: if $l_{th}$ represents the
(microscopic) scale of the dissipation,then a related scale $\sim
100 \times l_{th}$ exists. This is known as the {\it Markstein} length.
An inspection of eq.(1) shows
that the value of the exponent,however, does not allow a vary
large increase of the velocity. Actually the flame speeds up in
numerical simulations (~\cite{Ropke-Hillebrandt}) but not
dramatically. Independently of this, the cellular stage is short
and more dramatic effects happen along the propagation as
described below.

\section{Disruption of the flame and turbulent cascade}

\begin{figure}[htb]
\begin{center}
\epsfig{file=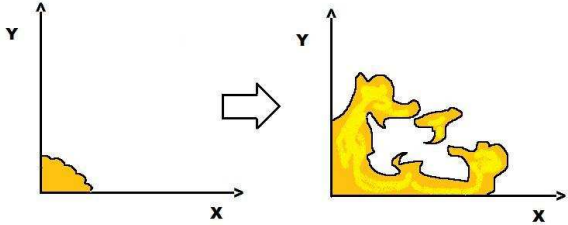,height=1.5in} \caption{Scheme of the
deformation of flame from the initial laminar stage to the
turbulent cascade (short cellular stage not shown).}
\label{fig:casc}
\end{center}
\end{figure}

The action of gravity directed against the propagation speed has
been observed to disrupt the cellular flame generating a {\it
turbulent cascade} also on short timescale. Above a length {\it L}
at which $u_{cell} = u^{'}(L)$ (where $u^{'}$ is the velocity
fluctuating part), disruption of bubbles occur. This is the
so-called {\it Gibson} scale. The estimation of this length
requires the specification of the turbulent spectrum. Assuming a
Kolmogorov form, one can check that

\begin{equation}
l_{Gibson} \propto \, {\big( {u_{cell}/u^{'}(L)}\big)}^{3} \, \leq
\, 10^{-4} cm
\end{equation}

since this is a small length, but still $\gg \, l_{th}$ because
the thermal scale is truly microscopic, we can classify the
propagation as belonging to the {\it flamelet regime}. In this
situation, the flame propagation is still controlled by diffusion,
but the total burning rate is determined by turbulence, and termed
the {\it flame brush}. When the flame brush is developed,
turbulent eddies turnover control the transport and fuel
consumption rate, thus $u_{turb}$ is now unrelated to diffusion. A
rough depiction of the evolution of the flame is given in
Fig~\ref{fig:casc}.

The determination of $u_{turb}$ is possible by resorting to the
basic expressions of the RT instability, although more elaborated
models can be found in the literature, for example the {\it
Sharp-Wheeler} \cite{SW} model, and also a fractal model quite similar to
the one employed for the cellular regime ~\cite{CG}. In the latter the
velocity of the flame is

\begin{equation}
u_{RT} = u_{lam} {\bigg({L\over{l_{min}}}\bigg)}^{D-2}
\end{equation}

with $D$ the fractal dimension of the front. For any reasonable
value of $D$, and given that $L$ is bounded from above by the
largest turbulent eddies and $l_{min}$ can not be smaller than the
Gibson scale, it is concluded that $u_{RT} \equiv u_{turb} \gg
u_{lam}$ deep inside the star, for distances $\leq 1 \, km$.

\section{Distributed regime and detonations}

The flame now accelerated to $u_{turb} \gg 10^{4} cm s^{-1}$ may
be still subject to changes according to the actual physical
conditions. One intriguing possibility is the disruption of the
flame by turbulence, a condition reached when $l_{Gibson} <
l_{th}$. In principle, this can be reached early along the
propagation, but it should be remembered that $l_{th}$ is
microscopic and therefore not easy to beat. However, the so-called
{\it distributed} regime, in which no definite flame front occurs,
is sometimes described by the mixing of fuel and ashes interacting
strongly with turbulent eddies.

Even though it is not clear that the distributed regime ensures,
it is important to consider its consequences, because observations
show that it is one of the preconditions for a jump to the {\it
detonation} branch. Alternatively, the burning may proceed
outwards with a high (but subsonic) velocity $u_{turb} \leq 10^{9}
cm s^{-1}$, much faster than suggested by the simplest laminar
analysis.

One of the popular proposed mechanisms for this jump to the
detonation branch is the so-called Zel'dovich gradient. It is
often described as a synchronic burning of a mixed (fuel+ashes)
region. As a necessary condition, the region should be small
($l_{c} \leq 10 cm$) ~\cite{H2}, and also the mixing time smaller
than the burning time, a condition written as

\begin{equation}
\tau_{mix} \leq {l_{c}\over{u'(l_{c})}}
\end{equation}

translated into a conservative bound $l_{c} < 10^{-5} cm$. This is
smaller, than the Gibson scale and makes the necessary condition
irrelevant. A thorough examination of this problem would require
the solution of the reactive Euler equations (see, for instance,
~\cite{euler}) and has not been attempted until now.

\section{Conclusions}

We have discussed semiqualitatively the realistic features of a $n
\rightarrow SQM$ burning in (just born) proto-neutron stars. The
conversion, which has been often treated as a simple phase
transition, should feature a quite complex evolution driven by LD
and RT instabilities, in a close analogy to thermonuclear
explosions of white dwarfs ~\cite{Klo}. These instabilities are
always present and act very promptly, this is why they can not be
ignored and make the laminar analysis just a ``zero-time''
academic exercise. After a few dynamical timescales the velocity
is likely to be high, but subsonic (turbulent deflagration), or
even supersonic (detonation) if the jump to that branch is
achieved. In both cases the outcome of the conversion should
affect the external layers, by blowing up and allow a bare quark
surface with high photon luminosity ~\cite{Xu} ~\cite{Usov}, by
enhanced neutrino heating onto a stalled prompt shock ~\cite{LB},
or by a direct piston action ~\cite{BH}. A small magnetic field is
also enough to produce substantial asymmetry of the front
~\cite{GLBH}, at least at a linear level. All these features
suggest that a new round of calculations is guaranteed.

\bigskip
I am grateful to FAPESP and CNPq Agencies, Brazil, for financial
support, and to the organizers of CSQCD II Workshop for
hospitality and support during this excellent event.

\end{document}